\documentclass[]{jfm}

\usepackage{graphicx}
\usepackage{caption}
\usepackage{subcaption}
\usepackage{subcaption}
\usepackage{newtxtext}
\usepackage{newtxmath}
\usepackage{natbib}
\usepackage{hyperref}

\hypersetup{
    colorlinks = true,
    urlcolor   = blue,
    citecolor  = black,
}

\newcommand{\RomanNumeralCaps}[1]
\linenumbers

\usepackage{cleveref}

\crefformat{section}{\S#2#1#3}

\title{Effect of rotation on turbulent mixing driven by the Faraday instability.}

\author{Narinder Singh\aff{1}
 and Anikesh Pal\aff{1}
  \corresp{\email{pala@iitk.ac.in}}}

\affiliation{\aff{1}Department of Mechanical Engineering, Indian Institute of Technology Kanpur, Kanpur 208016, U.P., India}

\begin{document}
\maketitle

\begin{abstract}
The effect of the rotation on the turbulent mixing of two miscible fluids of small contrasting density, produced by Faraday instability, is investigated using direct numerical simulations (DNS). We demonstrate that at lower forcing amplitudes, the turbulent kinetic energy ($t.k.e.$) increases with an increase in the Coriolis frequency $f$ till $\left(f/\omega\right)^2<0.25$, where $\omega$ is the forcing frequency, during the sub-harmonic instability phase. The increase in $t.k.e.$ increases the buoyancy flux ($B_V$), which increases the total potential energy (TPE). A portion of TPE is the available potential energy (APE). Some parts of APE  can convert to $t.k.e.$ via $B_V$, whereas the rest converts to internal energy, increasing background potential energy (BPE) through $\phi_i$. The remaining TPE also converts to BPE through the diapycnal flux $\phi_d$ resulting in irreversible mixing. With the saturation of the instability, irreversible mixing ceases. When  $\left(f/\omega\right)^2 > 0.25$, the Coriolis force significantly delays the onset of the sub-harmonic instabilities. During this period, the initial concentration profile diffuses to increase TPE, which eventually expends in BPE. The strong rotational effects suppress $t.k.e.$. Therefore, $B_V$ and APE become small, and the bulk of the TPE expends to BPE. Since the instability never saturates for $\left(f/\omega\right)^2 > 0.25$, the $B_V$ remains non-zero, resulting in a continuous increase in TPE. Conversion of TPE to BPE via $\phi_d$ continues, and we find prolonged irreversible mixing. At higher forcing amplitudes, the stabilizing effect of rotation is negligible, and the turbulence is less intense and short-lived. Therefore, the irreversible mixing phenomenon also ends quickly for $\left(f/\omega\right)^2<0.25$. However, when $\left(f/\omega\right)^2>0.25$ a continuous mixing is observed. We find that the turbulent mixing is efficient at lower forcing amplitudes and rotation rates of $\left(f/\omega\right)^2 > 0.25$. \\ 
 
\end{abstract}

\begin{keywords}
\end{keywords}
  

\section{Introduction} \label{sec:intro} 
The Faraday instability \citep{faraday1831xvii} describes the generation of standing wave patterns at the interface of two immiscible fluids subjected to vertical periodic vibration. However, in a stably stratified two-layer miscible fluid system, these standing waves become highly disorganized above a certain forcing amplitude and start to interact with each other resulting in the mixing of fluids \citep{zoueshtiagh2009experimental,amiroudine2012mixing,diwakar2015faraday}. \cite{grea2018final,briard2019harmonic} studied the onset and saturation of turbulent mixing owing to Faraday instability in fluids of small contrasting density for a wide range of parameters using theoretical models and DNS. Further experiments were carried out by \cite{briard2020turbulent,cavelier2022subcritical} using fresh and salty water to explore the dynamics of turbulent mixing. They found that when the instability is triggered, a natural wavelength appears at the interface between the two fluids. With the increase in amplitude, well-defined structures form that break to produce turbulent mixing. However, owing to the saturation of the instability, the turbulent mixing cannot be sustained for long. Recently, \citet{singh2022onset} presented a theoretical analysis of Faraday instability in miscible fluids under the effect of rotation and corroborated their findings with DNS. They reported that rotational effects stabilize the flow and delay the onset of the sub-harmonic instability at lower forcing amplitudes ($F$), whereas, with an increase in $F$, the stabilizing effect of rotation diminishes. They concluded that the instability saturates for $\left(f/\omega \right)^2<0.25$, and the mixing zone size asymptotes. However, for $\left(f/\omega \right)^2\geq0.25$, the instability does not saturate, and the mixing zone size continues to grow. The present investigation aims to quantify the turbulent mixing driven by Faraday instability in terms of the exchanges among the total potential energy (TPE), the background potential energy (BPE), and the available potential energy (APE) under the influence of rotation. \\   
\section{Problem formulation and numerical details}\label{sec:numerical details}
The two-layer miscible fluid system with small density contrast is considered here, which is driven by vertical periodic oscillations of acceleration $g(t)=g_0(1+F\cos{(\omega t))}$, where $g_0$ is the mean acceleration. We consider the density of the mixture to be a linearly varying function of mass concentration such that the lighter fluid with concentration $C(\rho_2)=0$ is placed above the denser fluid with $C(\rho_1)=1$ in a rectangular domain. The governing equations are the three-dimensional incompressible unsteady Navier–Stokes equations with Boussinesq approximation and are solved in a Cartesian coordinate system on a staggered grid arrangement as discussed in \citet{singh2022onset}. We consider periodic boundary conditions for all variables in the $x_1$ and $x_2$ (horizontal) directions, whereas at the top and bottom walls ($x_3$), no-slip boundary conditions for velocity vector $\boldsymbol{U}$ and Neumann boundary conditions for $C$ and pressure $P$ are used. The mixing zone size-$L$ is computed from mean concentration profile $\langle C \rangle$ \citep{andrews1990simple} as $L=6\int_{-\infty}^{+\infty} \langle C \rangle  (x_3,t) \left(1-\langle C \rangle (x_3,t)\right)\mathrm{d}x_3$, where $\langle \; \rangle$ denotes the horizontal average. The  Brunt V\"{a}is\"{a}l\"{a} frequency is defined as $N=\left({-2\mathcal{A} g_0{\partial \langle C \rangle}/{\partial x_3}}\right)^{1/2}$, where ${\partial \langle C \rangle}/{\partial x_3}=-1/L$. Table \ref{tab:parameters} shows the key parameters for all simulation cases. We select three cases at each $F$ depending on the value of $f/\omega$, where sub-harmonic instability saturates for $f/\omega=0$, and 0.48 ($\left(f/\omega \right)^2<0.25$) and never saturates for $f/\omega=0.59$ ($\left(f/\omega \right)^2\geq0.25$), as reported in \citet{singh2022onset}. We refer each case with a unique name for example F075f/$\omega$48, which indicates that $F=0.75$ and $f/\omega=0.48$.
\begin{table}
  \begin{center}
    \def~{\hphantom{0}}
    \setlength{\tabcolsep}{8pt} 
    \renewcommand{\arraystretch}{1.2} 
     \begin{tabular}{lcccccccc}
     $F$    &   $\omega \:(\mathrm{rad\:s^{-1}})$   &   $f/\omega$ ($\eta_{cu}\simeq$) &   $f/\omega$ ($\eta_{cu}\simeq$)  &   $f/\omega$ ($\eta_{cu}\simeq$)  \\[3pt]
    $0.75$  & 0.67  & 0 ($\simeq$0.52) & 0.48 ($\simeq$0.42)  &   0.59 (never saturate)  \\
    $1$     & 0.7   & 0 ($\simeq$0.44) & 0.48 ($\simeq$0.40)  &   0.59 (never saturate)   \\
    $2$     & 0.8   & 0 ($\simeq$0.41) & 0.48 ($\simeq$0.38)  &   0.59 (never saturate)   \\
    $3$     & 0.9   & 0 ($\simeq$0.355) & 0.48 ($\simeq$0.34)  &   0.59 (never saturate)   \\
  \end{tabular}
  \caption{Simulation parameters. For all cases we use Atwood number $\mathcal{A}=0.01$, initial mixing zone width $L_0=0.096 \:\mathrm{m}$, kinematic viscosity $\nu=1\times10^{-4}\:\mathrm{m^2\:s^{-1}} $, diffusion coefficient $\kappa=1\times10^{-4}\:\mathrm{m^2\:s^{-1}}$, and $g_0=10\:\mathrm{m\:s^{-2}}$ \citep{briard2019harmonic,singh2022onset}. Domain size: $l_{x_1}=l_{x_2}=2\pi \:\mathrm{m}$, and $l_{x_3}=2H=3.5\pi \:\mathrm{m}$. 
  Grid points: $N_{x_1}=N_{x_2}=512$ (uniform), and $N_{x_3}=512$ (non-uniform with clustering at the centre region of thickness $1.53\pi \:\mathrm{m}$ with $\Delta x_{3_{min}}=\Delta x_1=\Delta x_2$).
  }  \label{tab:parameters}
  \end{center}
\end{table}
\section{Results}\label{sec:results}
 \captionsetup[subfigure]{textfont=normalfont,singlelinecheck=off,justification=raggedright}
 \begin{figure}
 \centering
 \begin{subfigure}{0.425\textwidth}
	\centering
	\includegraphics[width=1.0\textwidth,trim={0cm 0.2cm 0cm 0cm},clip]{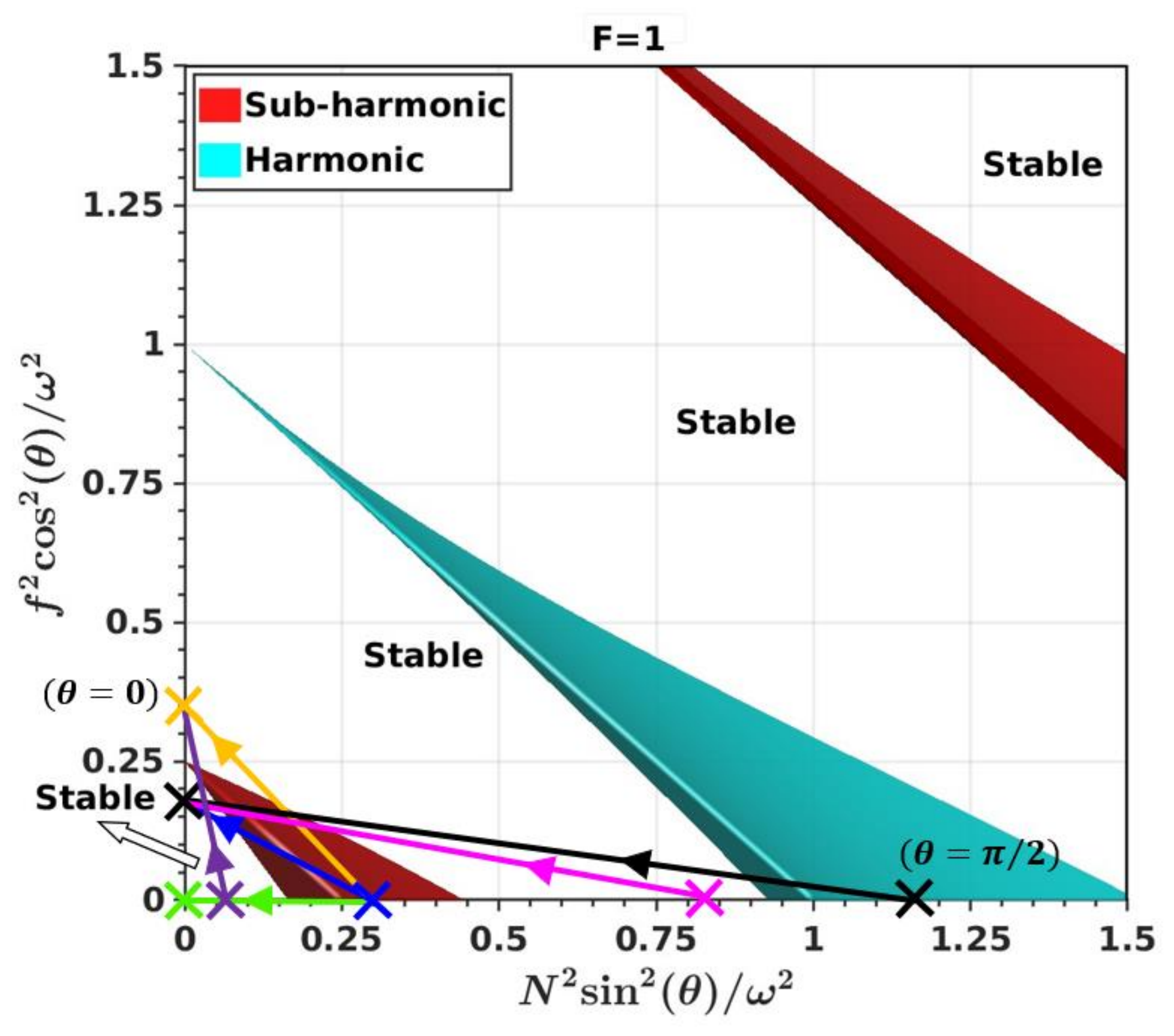}
	\caption{}	\label{subfig:3dF075}
 \end{subfigure}
 \quad
 \begin{subfigure}{0.425\textwidth}
	\centering
	\includegraphics[width=1.0\textwidth,trim={0cm 0.2cm 0cm 0cm},clip]{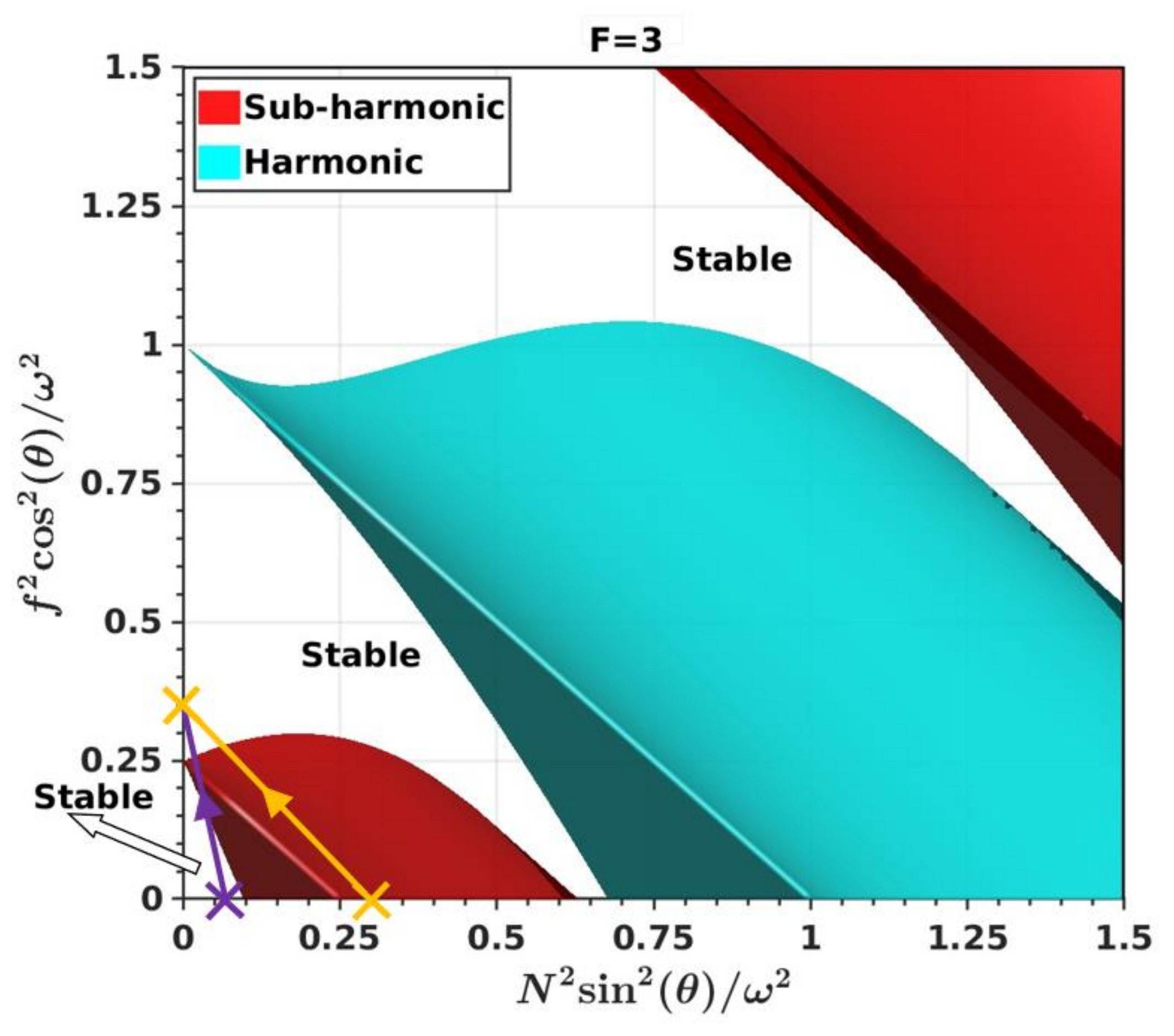}
	\caption{}	\label{subfig:3dF3}
 \end{subfigure}
 \caption{Top views of the three-dimensional Mathieu stability diagram \citep{singh2022onset} at forcing amplitudes (\textit{a}) $F=1$ and (\textit{b}) $F=3$. The stable (white) regions are in between the unstable red (sub-harmonic) and cyan (harmonic) colored tongues. The horizontal green line segment denotes the case without rotation ($f=0$) whereas rotation cases ($f\neq0$) are indicated by the inclined blue, pink, black, purple and orange line segments.}	\label{fig:3dFplane}
\end{figure}

\begin{figure}
 \centering 
  \begin{subfigure}{0.32\textwidth}
	\centering
	\includegraphics[width=1\textwidth,trim={0.1cm 0.1cm 0.1cm 0.1cm},clip]{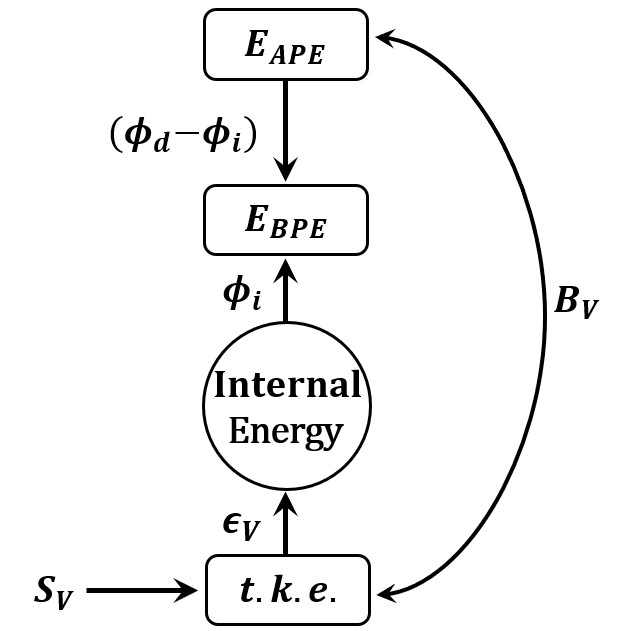}
	\caption{}  \label{subfig:BPE}
  \end{subfigure}
  \quad \quad
  \begin{subfigure}{0.554\textwidth}
	\centering
	\includegraphics[width=1\textwidth,trim={0.05cm 0.08cm 0.07cm 0.0cm},clip]{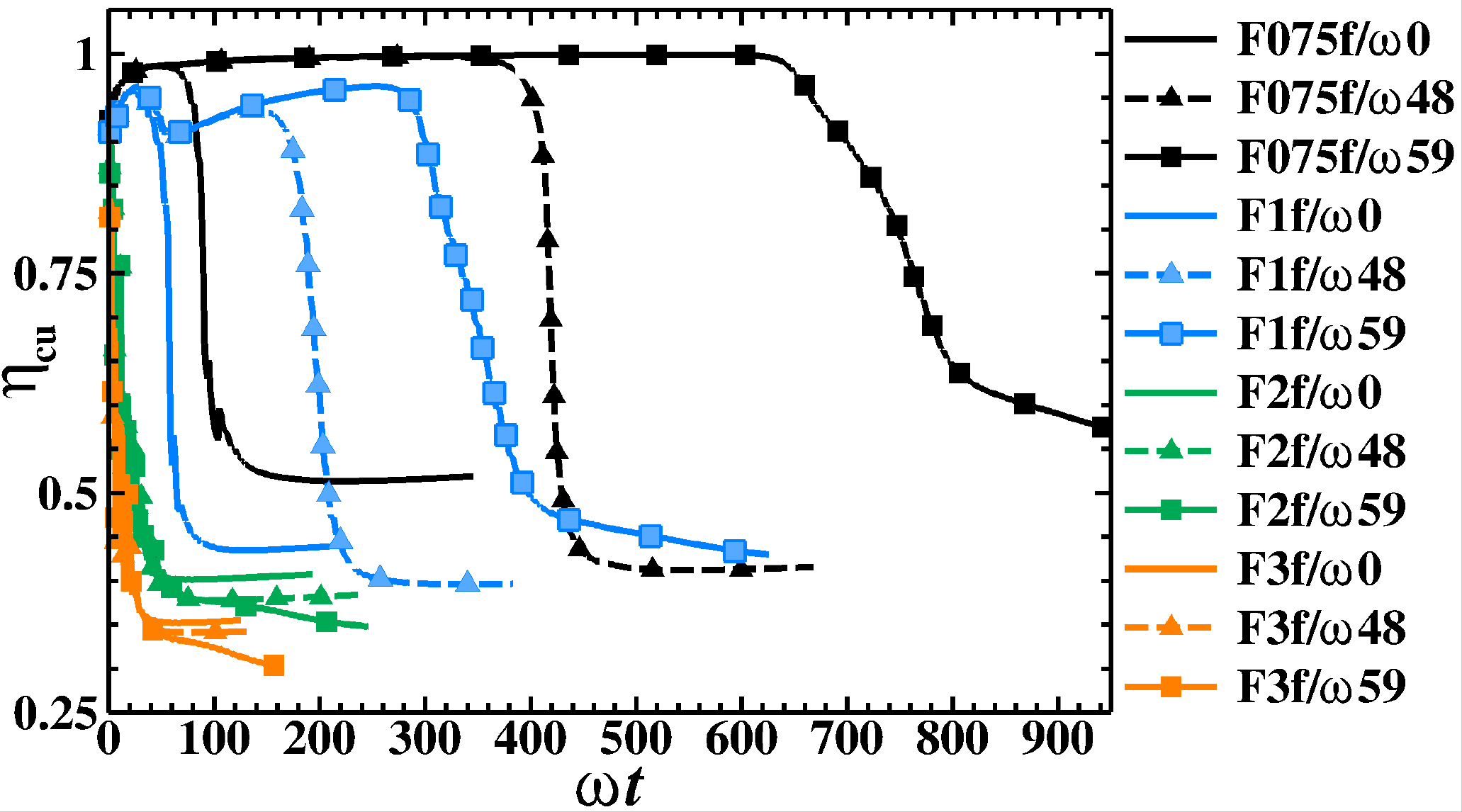}
	\caption{}  \label{subfig:eff}
  \end{subfigure}
  \caption {(\textit{a}) Schematic of energy pathways, $S_V$ is the volume integrated $S$. 
  (\textit{b}) time evolution of cumulative mixing efficiency $\eta_{cu}$; at forcing amplitudes $F=0.75$, 1, 2, and 3 for each without rotation ($f/\omega=0$) and with rotation ($f/\omega=0.48,\,0.59$) cases.}  \label{fig:PE eff}
 \end{figure}

We briefly explain the stability diagrams obtained by solving the linearized governing equations (Mathieu equations). \cite{singh2022onset} presents the details of the derivation of the linearized equations and their solution strategy. Figure \ref{fig:3dFplane} shows the stability diagrams at forcing amplitudes $F=1$ and $3$. In figure \ref{subfig:3dF075}, the horizontal green line segment represents the case without rotation $\left(f/\omega\right)^2=0$. The inclined blue, pink and black line segments represent the cases with $0<\left(f/\omega\right)^2<0.25$ whereas the purple and orange line segments denote the cases with $\left(f/\omega\right)^2\geq0.25$. The right end of each segment ($\times$; at $\theta=\pi/2$) corresponds to the mixing zone size-$L$ $\left( N^2/\omega^2=2\mathcal{A}g_0/\left(L \omega^2\right) \right)$, while the left end ($\times$; at $\theta=0$) corresponds to the Coriolis frequency $f$. When unstable $\theta$-modes in the sub-harmonic or harmonic tongues are excited by the periodic vertical forcing, $L$ grows, and $N^2/\omega^2$ moves toward the left. The $\theta$-modes are excited along the black line segment for $F = 1$  as shown in figure \ref{subfig:3dF075}. $L$ grows when the unstable $\theta$-modes along this line are excited. With time evolution, the right end of this inclined black line segment moves in the stable region between the leftmost sub-harmonic and harmonic tongues, as demonstrated by the pink $\times$. The process repeats, and $N^2/\omega^2$ moves into the leftmost sub-harmonic regime represented by the blue $\times$. The sub-harmonic instabilities are triggered in this regime resulting in turbulent mixing. In contrast to the horizontal green line segment, more unstable $\theta$-modes are excited along the inclined blue line segment resulting in the triggering of more instabilities and an increase in turbulent kinetic energy ($t.k.e.$) for cases with $0<\left(f/\omega\right)^2<0.25$ as compared to the case without rotation. The instability finally saturates for $\left(f/\omega\right)^2<0.25$ when no more unstable $\theta$-modes are excited, and the right end of the segment crosses the left boundary of the first sub-harmonic tongue and enters the leftmost stable region. For $\left(f/\omega\right)^2\geq0.25$, the orange line segment indicates that both unstable and stable $\theta$-modes are excited during the periodic vertical oscillations. The stable $\theta$-modes do not contribute to the triggering of instabilities, and therefore, we can deduce that higher rotation rates suppress turbulence. With an increase in $F$, the first sub-harmonic and harmonic tongues become wider, and the stable region between these tongues shrinks (figure \ref{subfig:3dF3}) resulting in the early onset of the sub-harmonic instability and turbulent mixing similar to the non-rotating case. An interesting feature for $\left(f/\omega\right)^2\geq0.25$ is observed when $N^2/\omega^2$ moves into the leftmost stable regime denoted by the purple $\times$ in figures \ref{subfig:3dF075} and \ref{subfig:3dF3}. The purple line segment must always pass through the first sub-harmonic tongue. Therefore, the instability will never saturate, resulting in the sustenance of turbulence. \\ 
\captionsetup[subfigure]{textfont=normalfont,singlelinecheck=off,justification=raggedright}
 \begin{figure}
 \hfill
 \begin{subfigure}{0.49\textwidth}
	\centering
	\includegraphics[width=1\textwidth,trim={0cm 0.25cm 0.1cm 0.0cm},clip]{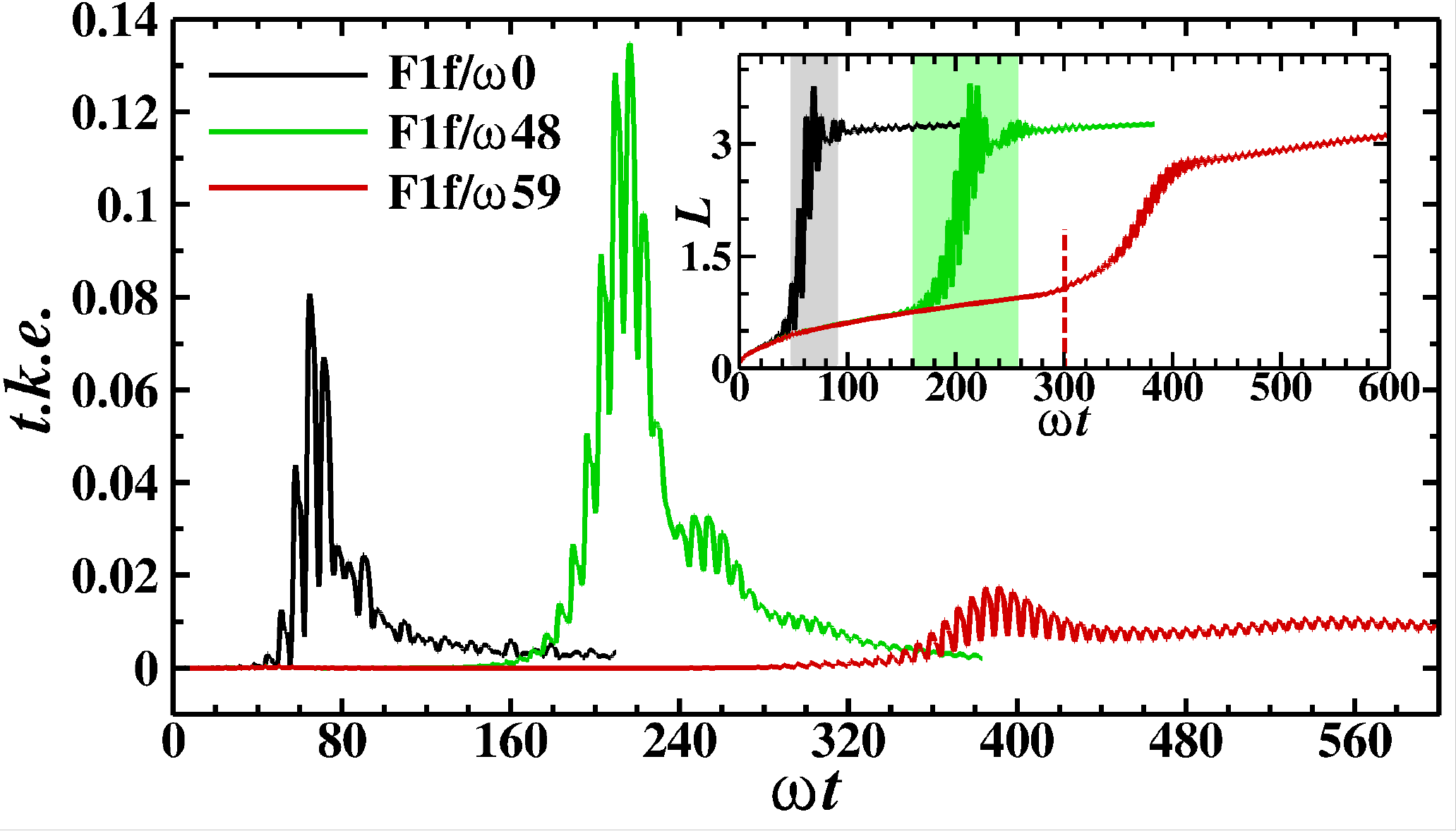}
	\caption{}	\label{subfig:tkeF1}
 \end{subfigure}
 \hfill
 \begin{subfigure}{0.49\textwidth}
	\centering
	\includegraphics[width=1\textwidth,trim={0cm 0.25cm 0.1cm 0.0cm},clip]{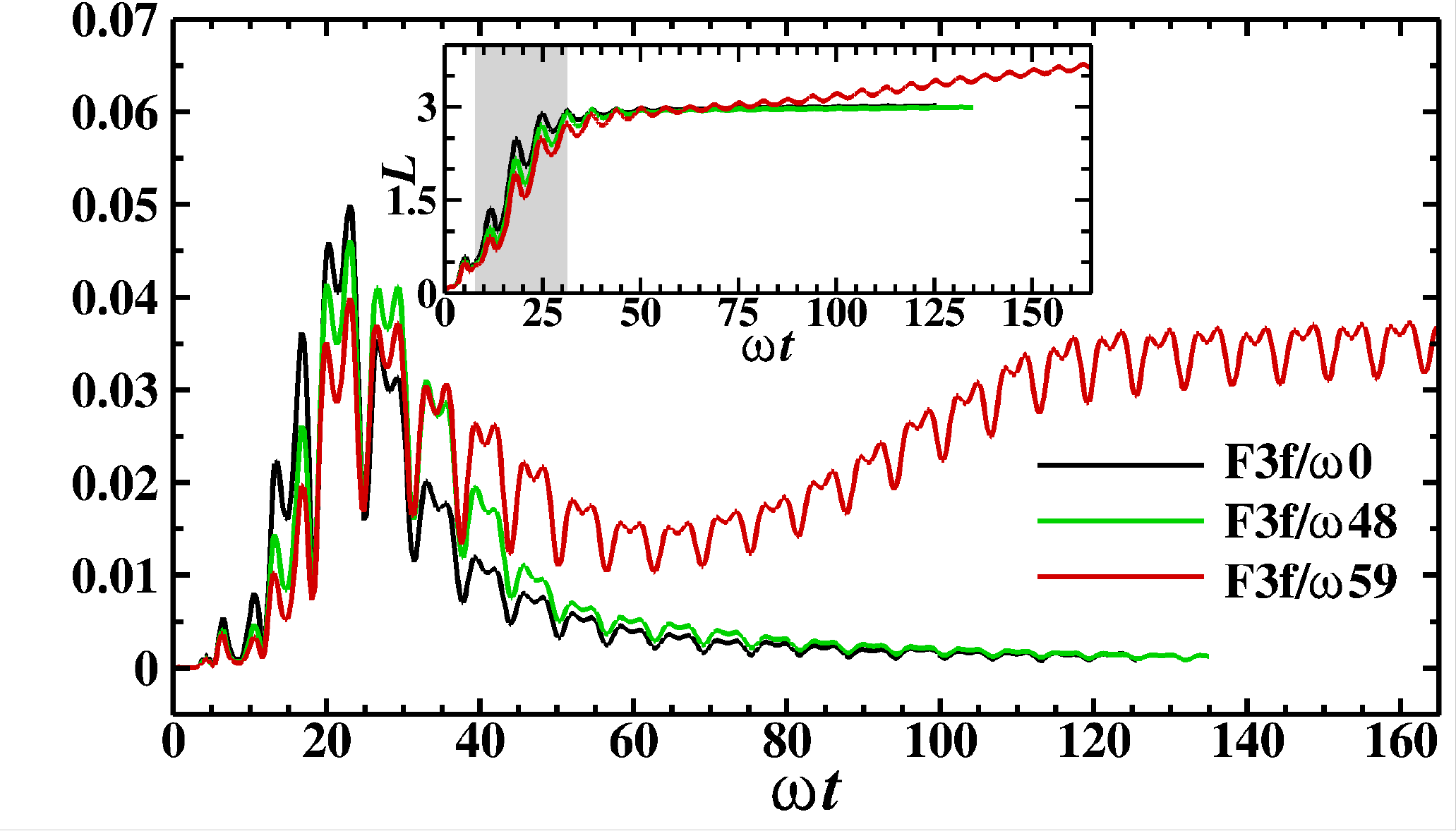}
	\caption{}	\label{subfig:tkeF3}
 \end{subfigure}
 \caption{Evolution of vertically ($x_3$) integrated $t.k.e.$ for both without ($f/\omega=0$) and with ($f/\omega=0.48,\,0.59$) rotation cases at (\textit{a}) $F=1$, and (\textit{b}) $F=3$. Inset shows the evolution of mixing zone size-$L$ with shaded regions indicating the sub-harmonic instability phase.} \label{fig:tke}
\end{figure} 
The evolution of vertically ($x_3$) integrated turbulent kinetic energy $t.k.e.$ ($(\langle u_1^2 \rangle +\langle u_2^2 \rangle + \langle u_3^2 \rangle) /2$, where $\boldsymbol{u}=\boldsymbol{U}-\langle \boldsymbol{U} \rangle$), for $F = 1$ %
is illustrated in figure \ref{subfig:tkeF1}. The $t.k.e.$ rapidly grows for F1f/$\omega$0 and F1f/$\omega$48 during the onset of the sub-harmonic instability. Eventually, the $t.k.e.$ decays owing to the instability saturation. The $t.k.e.$ for F1f/$\omega$48 is higher than F1f/$\omega$0 due to the excitement of more unstable $\theta$-modes in the sub-harmonic region, as discussed earlier. The $t.k.e.$ decreases significantly for $f/\omega=0.59$ owing to the suppression of turbulence with increased rotation rate. Although small, the $t.k.e.$ for F1f/$\omega$59 sustains owing to the continued triggering of the sub-harmonic instability, resulting in continuous turbulent mixing. A similar evolution of $t.k.e.$ is observed for $F = 0.75$. At a higher forcing amplitude of $F = 3$, the magnitude of $t.k.e.$ is similar for all $f/\omega$ cases (figure \ref{subfig:tkeF3}), indicating that the stabilizing effect of rotation is mitigated by the strong vertical forcing. Notice that the maximum value of $t.k.e.$ for  F1f/$\omega$48 is higher than F3f/$\omega$48. This observation is also true for the non-rotating cases at respective forcing amplitudes. The longer sub-harmonic instability phase for F1f/$\omega$48 is the reason for a higher value of $t.k.e.$ at F1f/$\omega$48 than at F3f/$\omega$48. The sub-harmonic instability phase is the time interval between the onset and the saturation of the sub-harmonic instability and is demonstrated using the shaded regions in the evolution of the mixing zone size $L$ in the inset. A long sub-harmonic instability phase for $F = 0.75, 1$ signifies a delay in the saturation of the instabilities triggered during the low amplitude oscillations. Therefore, the instabilities get sufficient time to evolve and break into turbulence. In contrast to $F = 0.75, 1$, at $F = 2, 3$, the sub-harmonic instability phase is short. The instabilities saturate quickly and therefore do not have enough time to evolve and intensify turbulence. Interestingly, both F2f/$\omega$59 (figure not shown) and F3f/$\omega$59 manifest a recovery in $t.k.e.$ owing to the continuous triggering of the sub-harmonic instabilities.\\
\captionsetup[subfigure]{textfont=normalfont,singlelinecheck=off,justification=raggedright}
 \begin{figure}
 \centering
 \begin{subfigure}{0.195\textwidth}
    \centering
     \includegraphics[width=1\textwidth,trim={0.1cm 0.1cm 0.1cm 0.0cm},clip]{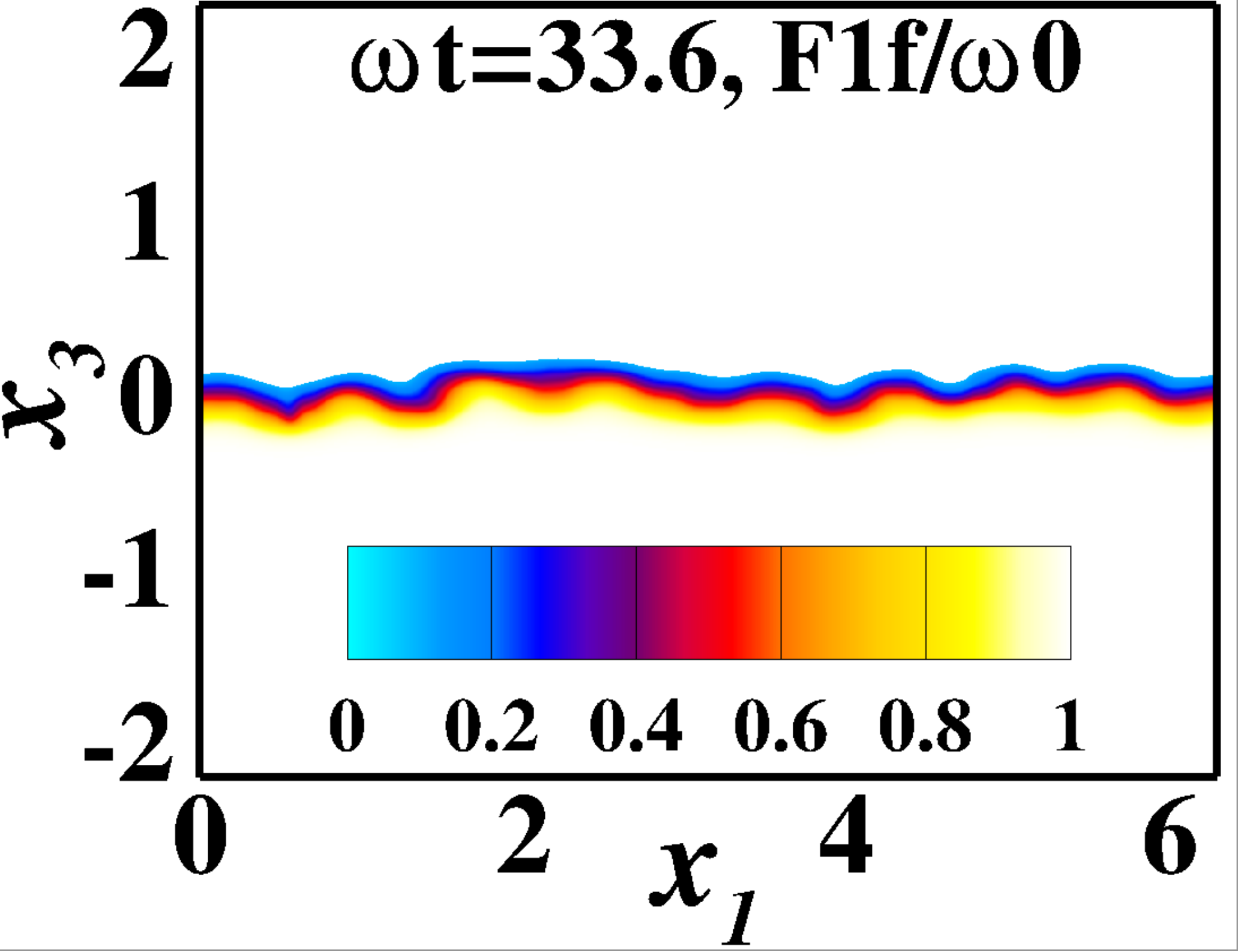}
    \caption{}  \label{subfig:C a}
 \end{subfigure}
 \begin{subfigure}{0.195\textwidth}
    \centering
    \includegraphics[width=1\textwidth,trim={0.5cm 0.1cm 0.1cm 0.0cm},clip]{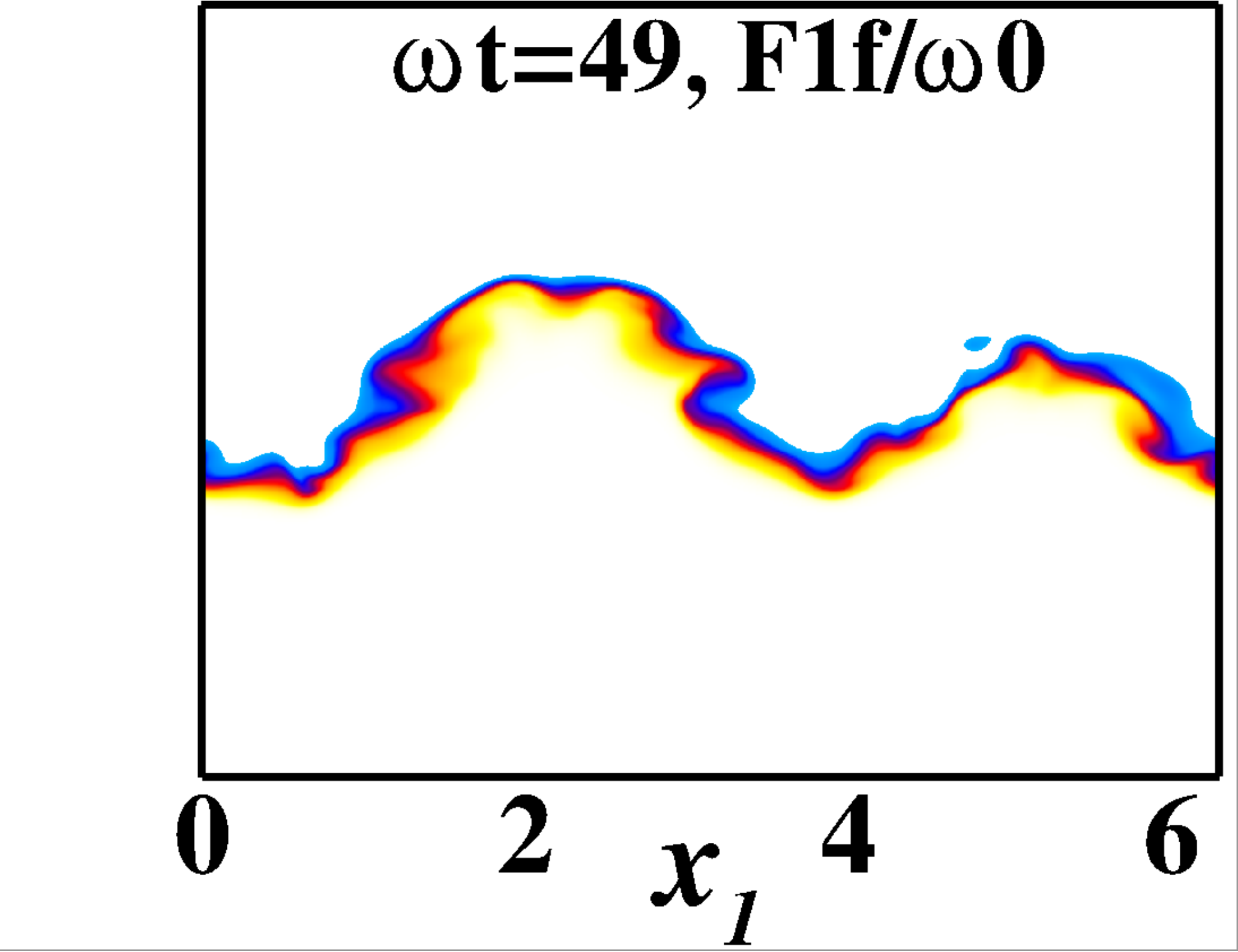}
    \caption{}  \label{subfig:C b}
 \end{subfigure}
 \begin{subfigure}{0.195\textwidth}
    \centering
    \includegraphics[width=1\textwidth,trim={0.5cm 0.1cm 0.1cm 0.0cm},clip]{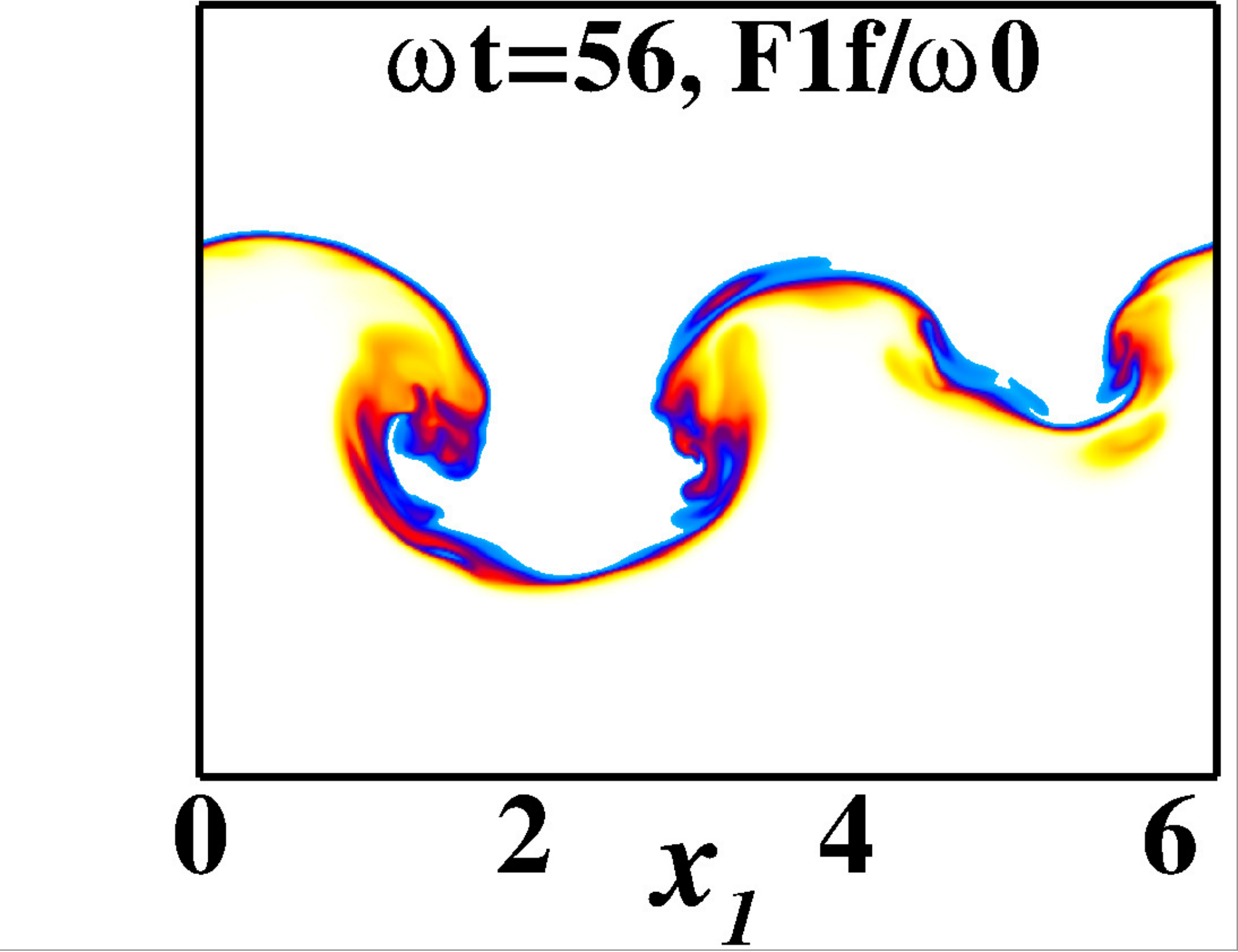}
    \caption{}  \label{subfig:C c}
 \end{subfigure}
 \begin{subfigure}{0.195\textwidth}
    \centering
    \includegraphics[width=1\textwidth,trim={0.5cm 0.1cm 0.1cm 0.0cm},clip]{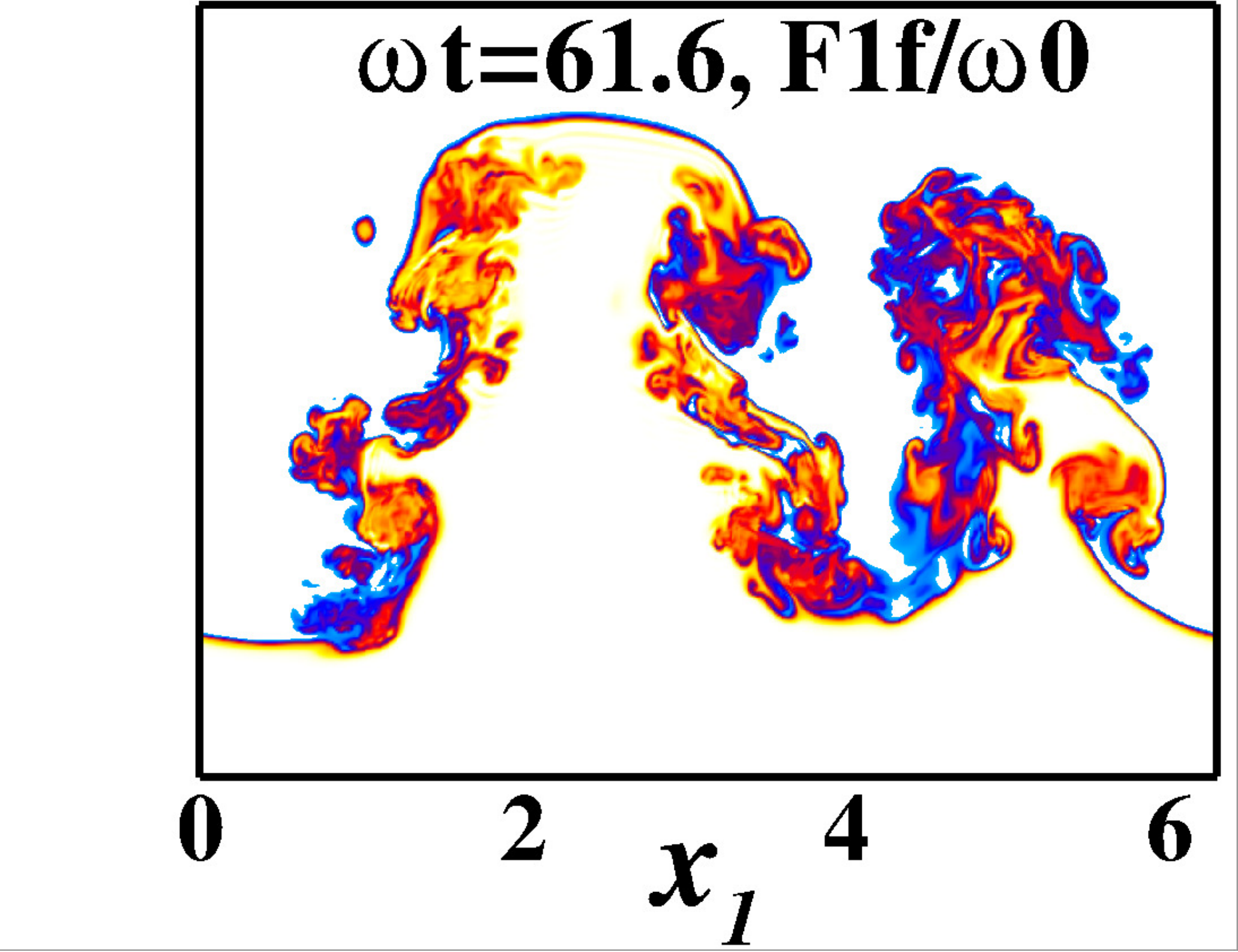}
    \caption{}  \label{subfig:C d}
 \end{subfigure}
 \begin{subfigure}{0.195\textwidth}
    \centering
    \includegraphics[width=1\textwidth,trim={0.5cm 0.1cm 0.1cm 0.0cm},clip]{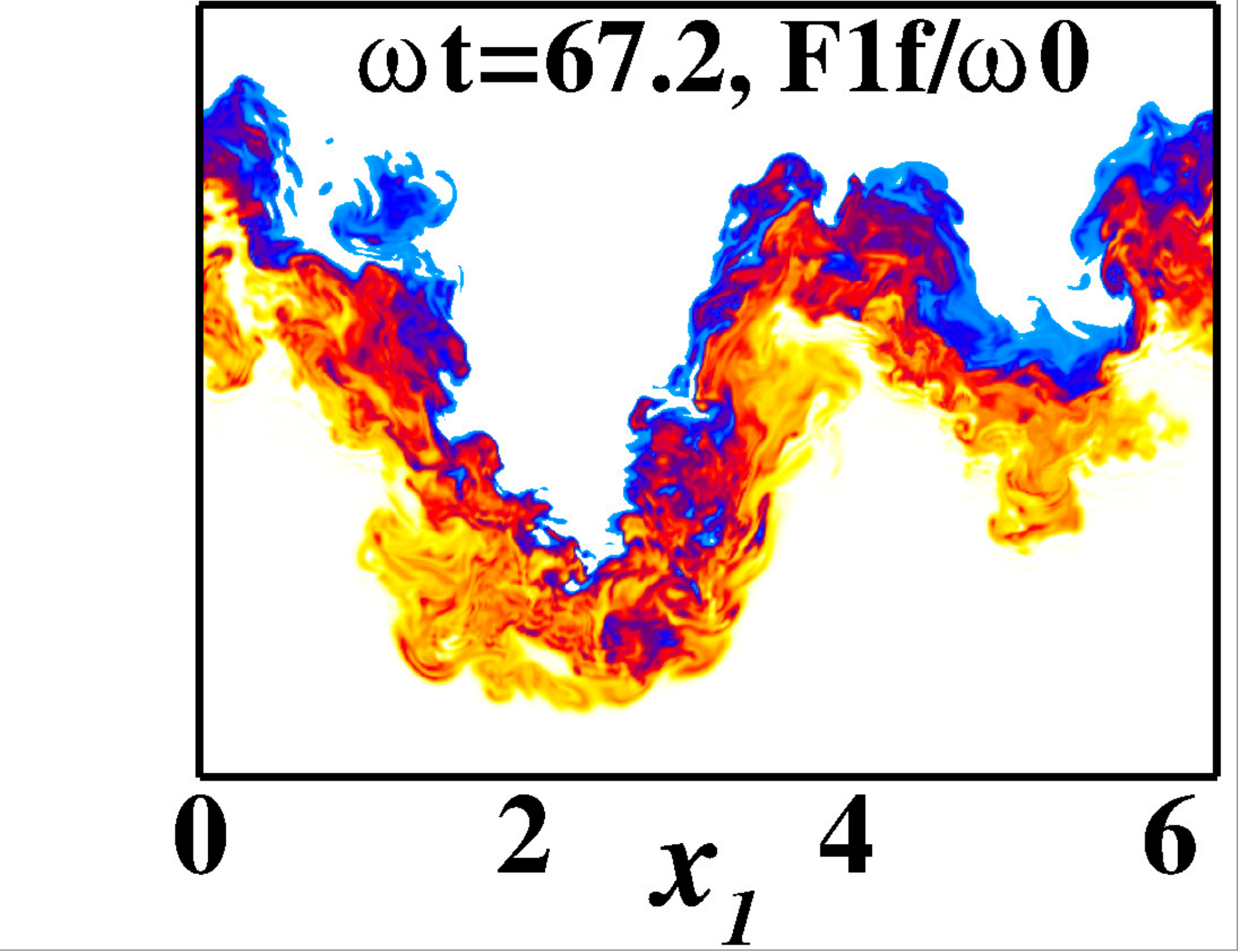}
    \caption{}  \label{subfig:C e}
 \end{subfigure}
 \caption{Contours of concentration field in vertical $x_1-x_3$ center plane ($x_2=\pi$) for case F1f/$\omega$0 at time instants (\textit{a}) $\omega t=33.6$, (\textit{b}) $\omega t=49$, (\textit{c}) $\omega t=56$, (\textit{d}) $\omega t=61.6$, and (\textit{e}) $\omega t=67.2$. Pure fluids of $C=1$ (denser) and $C=0$ (lighter) are made transparent.}   \label{fig:mean conc}
 \end{figure}
\captionsetup[subfigure]{textfont=normalfont,singlelinecheck=off,justification=raggedright}
 \begin{figure}
 \centering
 \begin{subfigure}{0.325\textwidth}
    \centering
    \includegraphics[width=1\textwidth,trim={0.1cm 0.3cm 0.6cm 0.2cm},clip]{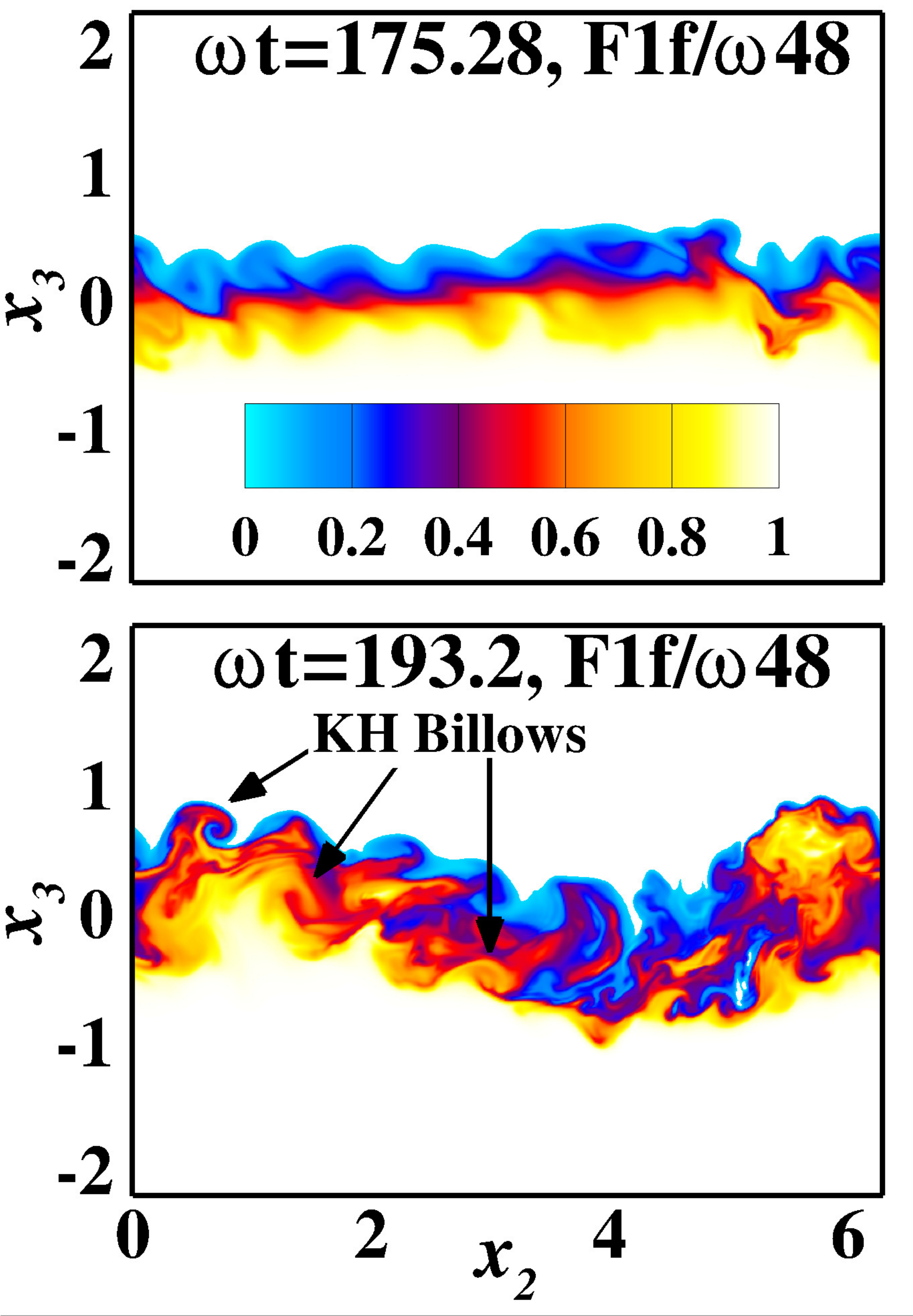}
    \caption{}  \label{subfig:C F1fw48}
 \end{subfigure}
 \begin{subfigure}{0.325\textwidth}
    \centering
    \includegraphics[width=1\textwidth,trim={0.1cm 0.3cm 0.6cm 0.2cm},clip]{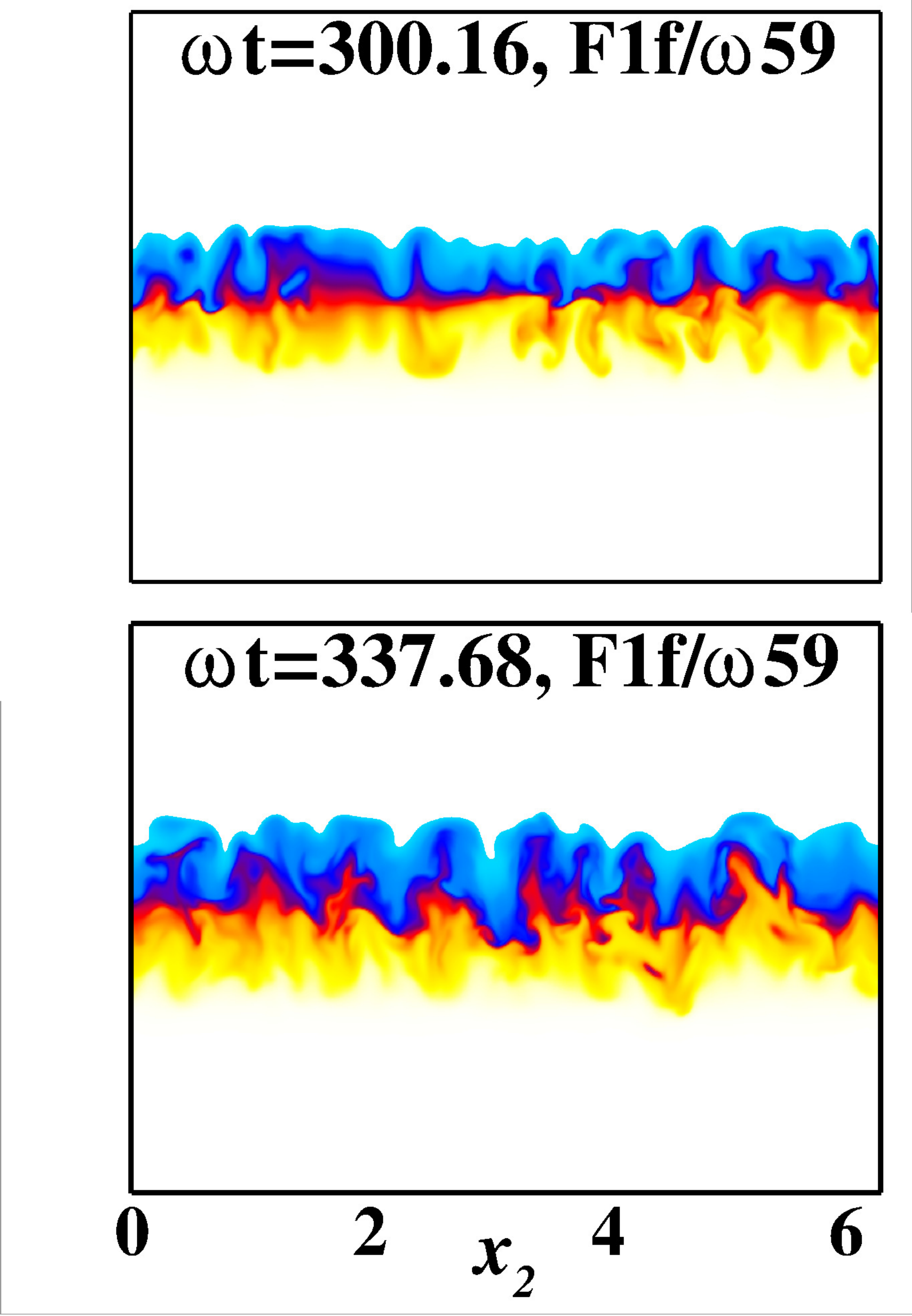}
    \caption{}  \label{subfig:C F1fw59}
 \end{subfigure}
 \begin{subfigure}{0.325\textwidth}
    \centering
    \includegraphics[width=1\textwidth,trim={0.1cm 0.3cm 0.6cm 0.2cm},clip]{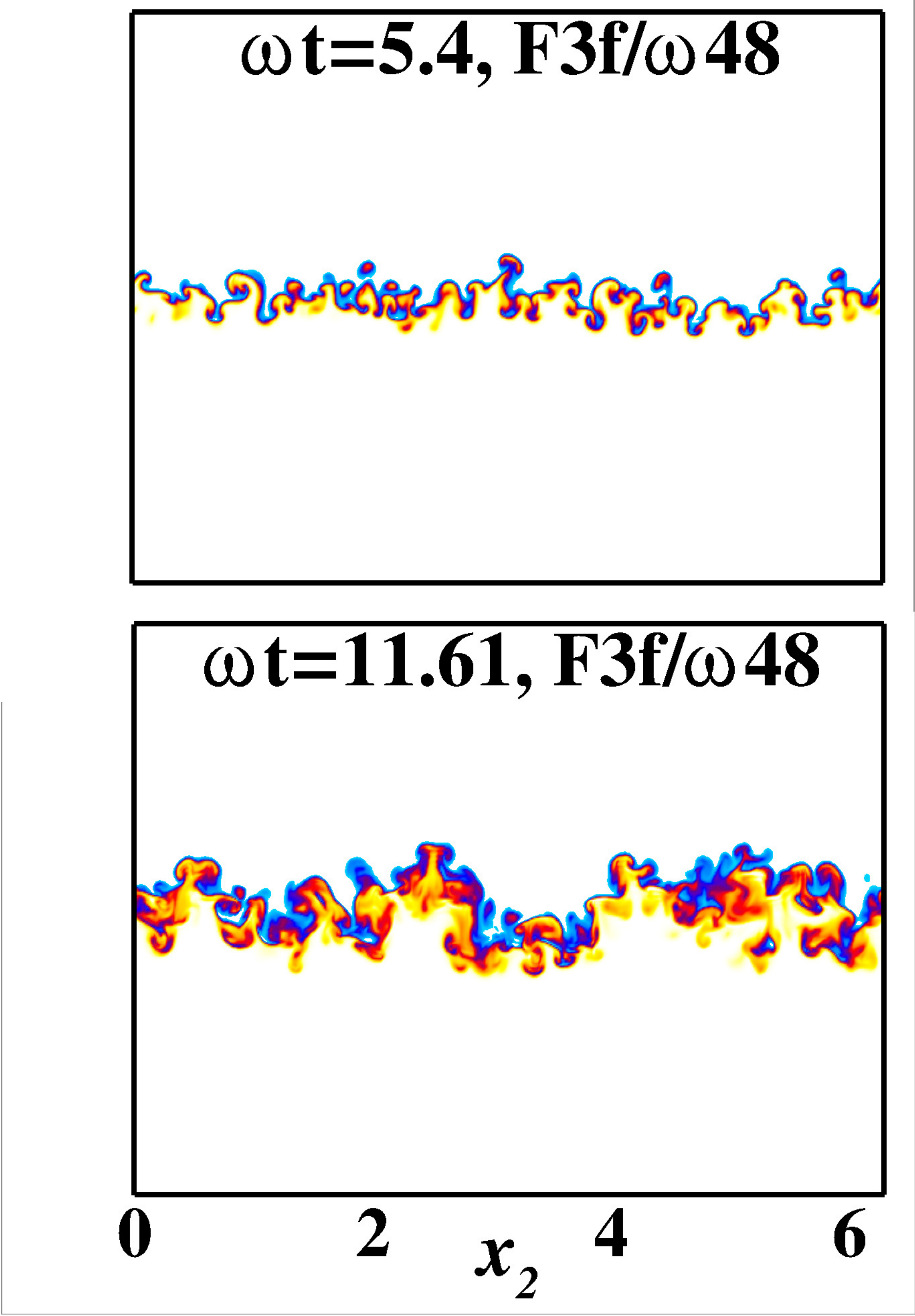}
    \caption{}  \label{subfig:C F3fw48}
 \end{subfigure}
 \caption{Contours of concentration field in vertical $x_2-x_3$ center plane ($x_1=\pi$) for rotation cases (\textit{a}) F1f/$\omega$48, (\textit{b}) F1f/$\omega$59, and (\textit{c}) F3f/$\omega$48 at different time instants. 
 }    \label{fig:conc}
 \end{figure}
The transition to turbulence phenomenon for the cases without and with rotation is analyzed using the contour plots of the concentration fields in a vertical plane. Figure \ref{fig:mean conc} depicts the evolution of the interface of the concentration field for F1f/$\omega$0. The initially perturbed interface, subjected to vertical and periodic accelerations, deforms into a wavy form owing to harmonic and sub-harmonic resonance as shown in figure \ref{subfig:C a} at $\omega t=33.6$. This diffused wavy interface is amplified by the sub-harmonic instability, leading to the formation of mushroom-shaped waves as shown in figure \ref{subfig:C b} at $\omega t=49$. In this instance, some waves with shorter wavelengths become disorganized and interact with each other, resulting in  small velocity fluctuations. In the next oscillation, the amplitude of a longer wavelength wave amplifies, leading to the birth of a well-defined mushroom-shaped wave that bursts at the nodes 
as depicted in figure \ref{subfig:C c} at $\omega t=56$. This process is known as wave-breaking of the Faraday waves \citep{cavelier2022subcritical} and is responsible for the onset of irreversible mixing. In the successive oscillations, the entire interface breaks down into small-scale structures leading to a full transition to turbulence as manifested in figures \ref{subfig:C d} and \ref{subfig:C e}. A similar dynamics is observed for $F = 0.75$ and $f/\omega=0$ (see Movie $1$ and Movie $4$).\\
The combined effect of vertical oscillation and rotation in F1f/$\omega48$ results in roll-ups at multiple locations on the diffuse interface as shown in figure \ref{subfig:C F1fw48} at $\omega t=175.28$. In subsequent oscillations, roll-ups will continue to form with a large size but in the opposite direction of the previous one. These roll-ups will finally take the form of Kelvin-Helmholtz (KH) billows that will break into turbulence (see figure \ref{subfig:C F1fw48} at $\omega t=193.2$). Notice that the entire interface layer breaks into small structures in one oscillation resulting in a larger thickness of the turbulent mixing region as compared to F1f/$\omega0$ where the mushroom-shaped wave first breaks at the nodes and then turbulence spreads to the entire interface layer in the subsequent oscillations. Movie $2$ included as the supplementary data demonstrates this event. This phenomenon is also responsible for larger $t.k.e.$ in F1f/$\omega48$ as compared to F1f/$\omega0$ (figure \ref{subfig:tkeF1}). We observe similar dynamics for F075f/$\omega48$ (see Movie $5$). This phenomenon results in a gradual increase of irreversible mixing. When the Coriolis frequency increase to $f/\omega=0.59$ (case F1f/$\omega59$), we found that the effect of rotation is strong enough to suppress the formation of large amplitude waves and KH billows. Therefore, the turbulent mixing is only caused by the finger-shaped instability as depicted in figure \ref{subfig:C F1fw59} (see supplementary Movie $3$ and Movie $6$).\\  
At higher forcing amplitude $F=3$ for $f/\omega=0.48$, we observe the development of random finger-shaped structures on the interface that break in the first oscillation as shown in the figure \ref{subfig:C F3fw48}. Therefore, the turbulent mixing starts from the beginning of the periodic forcing. Additionally, the intensity of turbulence is low because of the shorter sub-harmonic instability phase at higher $F$. As the stabilizing effect of rotation is subdued at higher forcing amplitudes \citep{singh2022onset}, the turbulent mixing occurs by this mechanism for all $f/\omega$ at $F=2,3$ (figures not shown). We have included the animation of the concentration field for F3f/$\omega48$ as supplementary data (Movie $7$).\\
Since there is no mean velocity field involved, we start the analysis of the irreversible mixing using the $t.k.e.$ evolution equation as follows: 
\begin{equation}
\label{tke budget}
 \frac{dt.k.e.}{dt}=\mathcal{P}-S-B-\epsilon-\frac{\partial \mathcal{T}_j}{\partial x_j}.
\end{equation}
The turbulent production term $\mathcal{P}=- \langle u_i u_j \rangle  \frac{\partial \langle U_i \rangle}{ \partial x_j}$ and the transport term $\partial\mathcal{T}_j/\partial x_j$, where $\mathcal{T}_j=\langle p u_j \rangle - 2\nu  \bigl \langle s_{ij} u_i \bigr \rangle + \frac{1}{2}\langle u_i u_i u_j \rangle$ are negligible. The energy input from the periodic forcing is $S=2\mathcal{A}g_0 F\cos{(\omega t)} \langle u_3 c \rangle$. The buoyancy flux $B=2\mathcal{A}g_0 \langle u_3 c \rangle$ accounts for the reversible rate of exchange between $t.k.e.$ and the total potential energy. The viscous dissipation $\epsilon=2\nu \bigl \langle s_{ij} s_{ij} \rangle$, where $s_{ij}=\left( \partial u_i / \partial x_j + \partial u_j / \partial x_i\right)/2$, acts as a sink for $t.k.e.$. We obtain an excellent closure of equation \ref{tke budget} for all the simulations signifying sufficient grid spacing in all directions to resolve all length scales.\\ 
The irreversible mixing is characterized by partitioning the total potential energy (TPE) $E_{TPE}= 2 \mathcal{A}g_0\int_{V} C(\boldsymbol{x}) x_3 \mathrm{d}V$ into the sum of an available potential energy (APE) $E_{APE}$ and background potential energy (BPE) $E_{BPE}$ \citep{winters1995available}. The BPE is the minimum potential energy a flow can have and is unavailable for extraction from the potential energy reservoir for driving macroscopic fluid motion. To estimate BPE of the flow at any instance of time, a volume element $dV$ with concentration $C$ at height $x_3$ is adiabatically relocated to $x_3^*$ such that the redistributed  concentration field is statistically stable ($\mathrm{d}C(x_3^*)/\mathrm{d}x_3^* \leq0$) everywhere. We compute $E_{BPE}= 2 \mathcal{A}g_0\int_{V} C(x_3^*) x_3^* \mathrm{d}V$ and subtract it from $E_{TPE}$ to obtain $E_{APE}$ \citep{winters2013available,chalamalla2015mixing}. Since our domain is closed by the top and bottom boundaries, the surface fluxes are zero. We derieve the simplified equations for the evolution of TPE, BPE, and APE following \cite{winters1995available},
 \begin{subequations}
    \begin{equation}
     \label{dTPEdt}
     \frac{dE_{TPE}}{dt}= B_V+\phi_i,
    \end{equation}
    \begin{equation}
     \label{dBPEdt}
     \frac{dE_{BPE}}{dt}= \phi_d,
    \end{equation}
    \begin{equation}
     \label{dAPEdt}
     \frac{dE_{APE}}{dt}= B_V-\left(\phi_d - \phi_i \right).
    \end{equation}
\end{subequations}
Here $B_V$, $\phi_i$, and $\phi_d$ are integrated over the volume $V=l_{x_1} l_{x_2} H_s$, and $H_s$ is the vertical height of the domain excluding top and bottom sponge layer thickness. The $B_V=\int{B l_{x_1} l_{x_2} \mathrm{d}x_3}$ is volume integrated buoyancy flux and $\phi_i=-2 \mathcal{A}g_0 \kappa A \left(\langle C_{top} \rangle-\langle C_{bot} \rangle\right)$, where $A=l_{x_1} l_{x_2}$, the irreversible rate of conversion of internal to potential energy. $\phi_d=2 \mathcal{A}g_0 \kappa \int_{V} -\frac{\mathrm{d}x_3^*}{\mathrm{d}C} |\nabla C|^2 \mathrm{d}V$ represents the rate of change of BPE due to irreversible diapycnal flux and is a measure of the irreversible mixing. Notice that as $\phi_d \geq0$, it always acts as a sink of APE, increasing $E_{BPE}$. Since we have periodic forcing in the domain, we choose to calculate the cumulative mixing efficiency \citep{peltier2003mixing,briard2019harmonic} as follows:
\begin{equation}
 \label{efficiency}
    \eta_{cu}(t)=\frac{\int_{t1}^{t2}\phi_d \;\mathrm{d}t}{\int_{t1}^{t2}\left(\phi_d +\epsilon_V\right)\mathrm{d}t}.
 \end{equation}
 Here $\epsilon_V=\int{\epsilon l_{x_1} l_{x_2} \mathrm{d}x_3}$ is the volume integrated viscous dissipation. The cumulative mixing efficiency $\eta_{cu}$ measures the irreversible loss of initially available energy that expends in irreversible mixing in contrast to the total energy loss to the irreversible mixing and viscous dissipation.\\
 \captionsetup[subfigure]{textfont=normalfont,singlelinecheck=off,justification=raggedright}
  \begin{figure}
 	\centering
  	\begin{subfigure}{0.48\textwidth}
  		\centering
  		\includegraphics[width=1.0\textwidth,trim={0cm 1.2cm 0.07cm 0cm},clip]{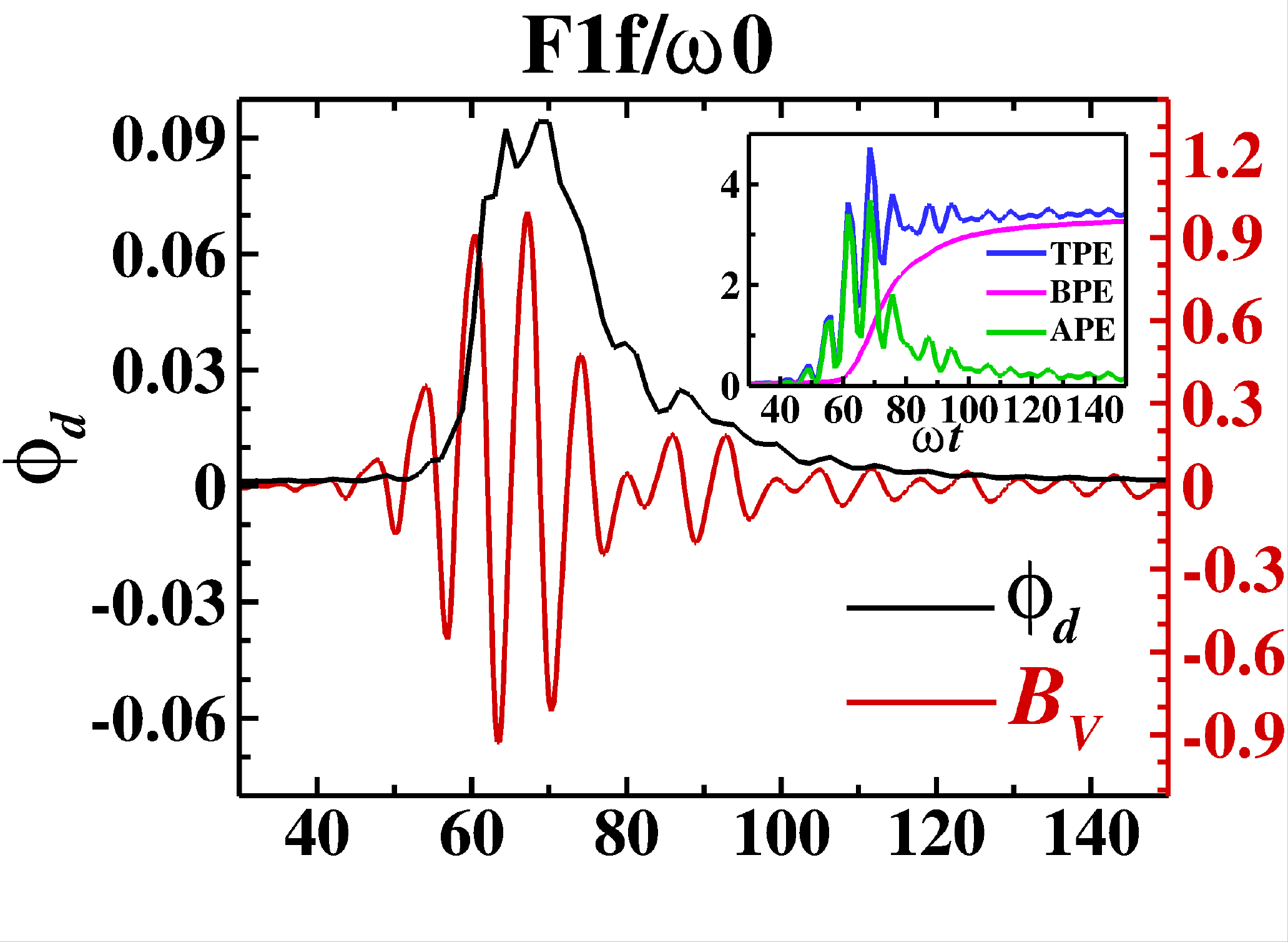}
   		\caption{}  \label{subfig:F1fw0}
  	\end{subfigure}
 	\quad 
   	\begin{subfigure}{0.48\textwidth}
   		\centering
   		\includegraphics[width=1.0\textwidth,trim={0cm 1.2cm 0.07cm 0cm},clip]{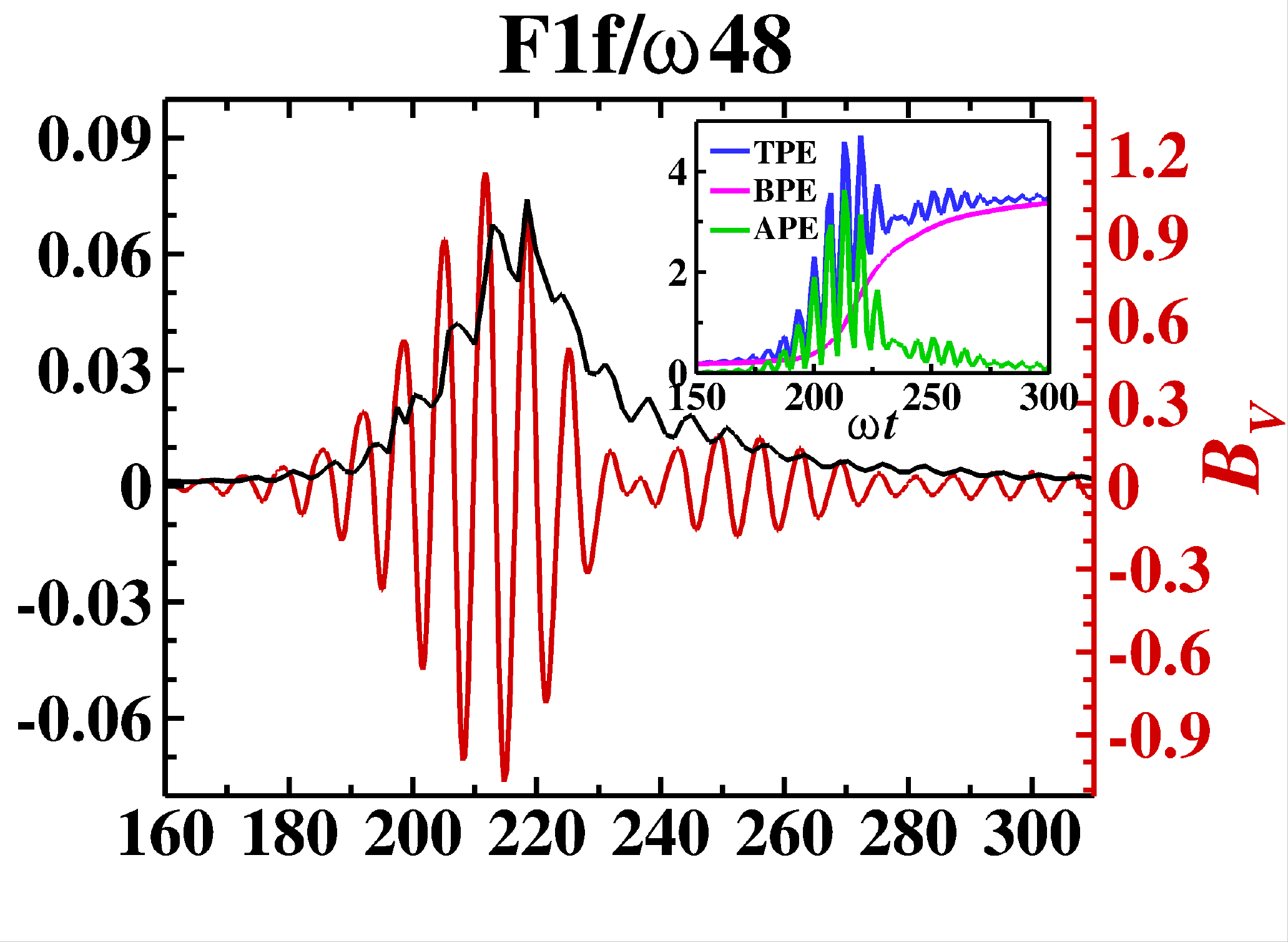}
   		\caption{}  \label{subfig:F1fw48}
   	\end{subfigure}
   	\begin{subfigure}{0.48\textwidth}
  		\centering
   		\includegraphics[width=1.0\textwidth,trim={0cm 0.4cm 0.07cm 0cm},clip]{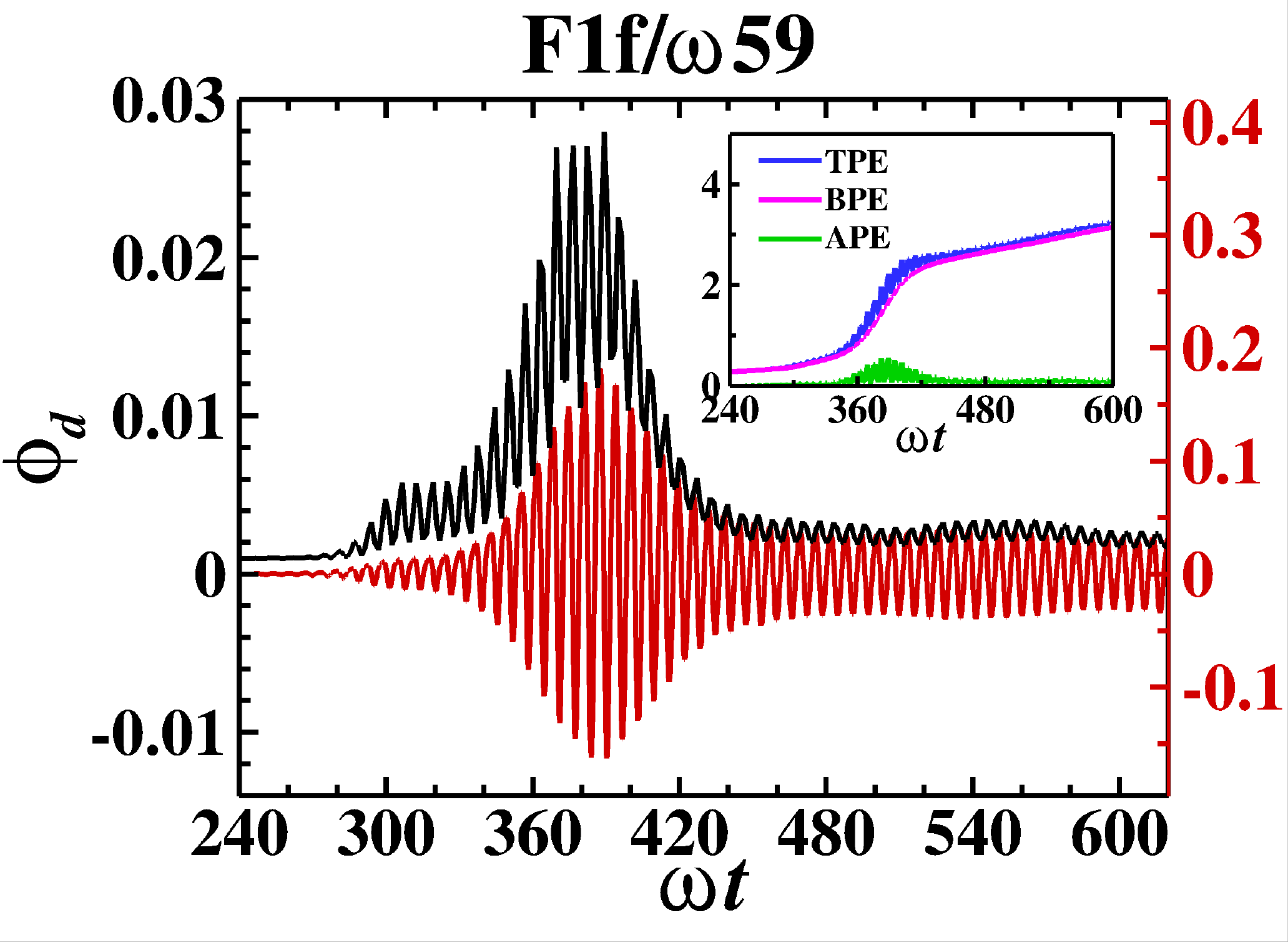}
   		\caption{}  \label{subfig:F1fw59}
   	\end{subfigure}
   	\quad
   	\begin{subfigure}{0.48\textwidth}
   		\centering
   		\includegraphics[width=1.0\textwidth,trim={0cm 0.4cm 0.07cm 0cm},clip]{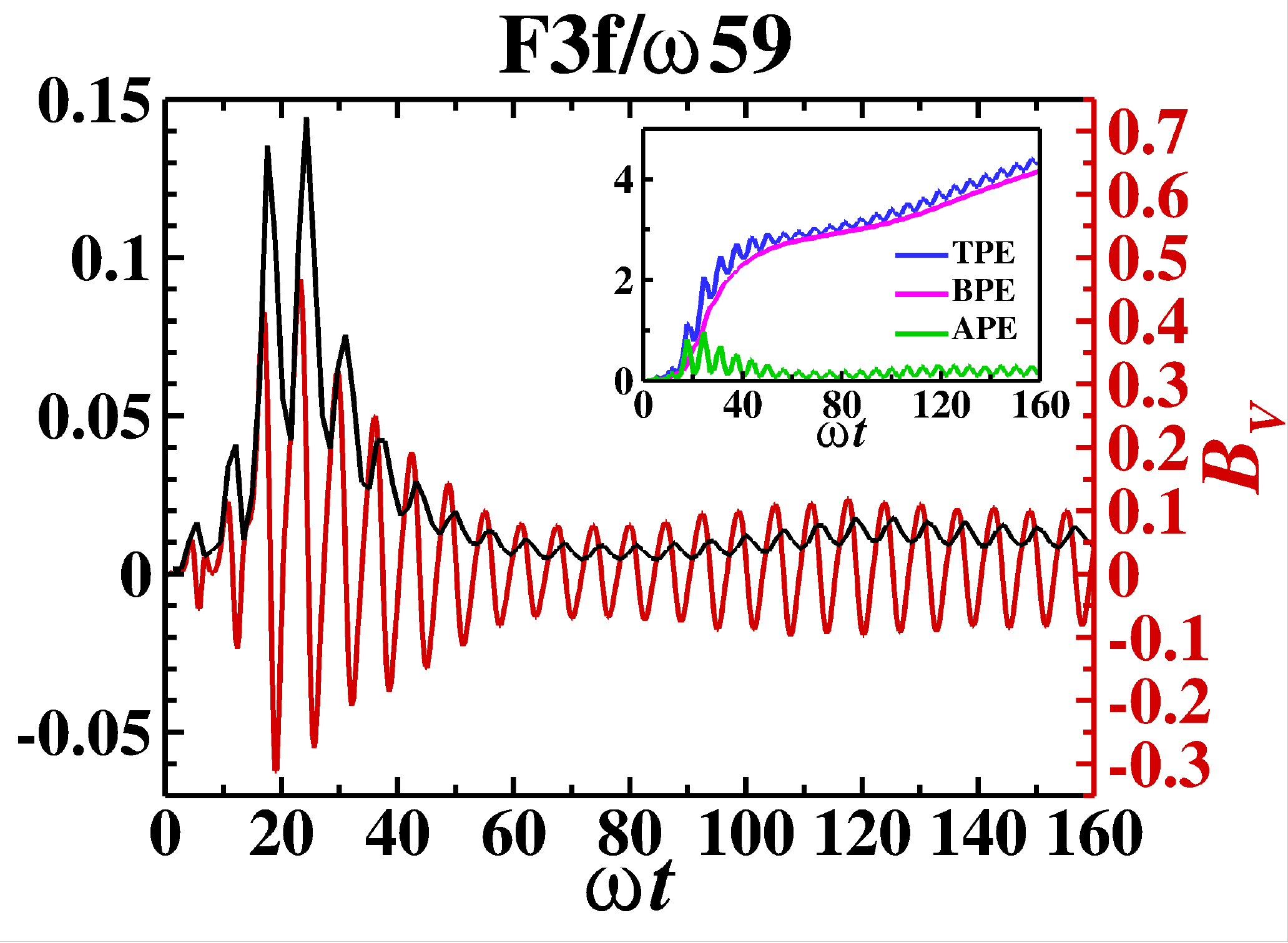}
   		\caption{}  \label{subfig:F3fw59}
   	\end{subfigure}
  	\caption{Time evolution of buoyancy flux ($B_V$) 
   and diapycnal flux ($\phi_d$) at $F=1$ for (\textit{a}) $f/\omega=0$, (\textit{b}) $f/\omega=0.48$, (\textit{c}) $f/\omega=0.59$, and at $F=3$ for (\textit{d}) $f/\omega=0.59$. Insets: correspond to the evolution of TPE, BPE, and APE.}    \label{fig:energy rates}
   \end{figure}
Figure $5$ illustrates the temporal evolution of the reversible buoyancy flux $B_V$ and the diapycnal flux $\phi_d$. We also include the evolution of TPE, APE, and BPE in the inset of each figure. The $t.k.e.$ for F1f/$\omega$0 increases due to the onset of sub-harmonic instabilities and the formation and breaking of mushroom-shaped waves. This increased $t.k.e.$ results in an increase in $B_V$, and therefore, TPE also increases as demonstrated in figure \ref{subfig:F1fw0}. A fraction of this TPE is APE. Some part of this APE can convert back to $t.k.e.$ via $B_V$ and the rest can convert to internal energy that increases BPE owing to $\phi_i$. Since $B_V$ increases, APE also increases at $\omega t\simeq46-55$. The remaining portion of TPE goes to BPE via $\phi_d$. The $\phi_d$ also starts increasing after $\omega t \simeq55$. After wave-breaking, the entire interface layer breaks, resulting in a rapid increase of $\phi_d$ signifying rapid conversion of TPE into BPE. When the instabilities saturate at $\omega t \simeq 70$, $t.k.e.$, $B_V$ and APE decrease, and TPE saturates. The BPE also saturates, as demonstrated by a decreasing $\phi_d$. Similar to F1f/$\omega$0, we observe that $B_V$ for F1f/$\omega$48 increases at $\omega t\simeq180$ owing to an increase in $t.k.e.$ (see figure \ref{subfig:F1fw48}). At the same instance, $B_V$, TPE, and APE also increase. We find a gradual increase of $\phi_d$ and BPE for F1f/$\omega$48 compared to a rapid increase for F1f/$\omega$0. The reason is the continuous formation and breaking of the KH billows at the diffuse interface during the sub-harmonic instability phase for F1f/$\omega$48. With the saturation of the instability, $B_V$ and APE start decreasing, resulting in the increase and eventual saturation of BPE as illustrated by the gradual decrease of $\phi_d$ to zero in figure \ref{subfig:F1fw48}. This gradual decrease in $\phi_d$ denotes that irreversible mixing is sustained for an extended period than F1f/$\omega$0, owing to the long sub-harmonic instability phase. For F1f/$\omega$59, there is a significant delay in the onset of sub-harmonic instability. Therefore, $B_V$ and APE $\sim 0$ till $\omega t \simeq 280$. During this period, the initial concentration profile diffuses \citep{winters1995available}, increasing the TPE at the rate $\phi_i$. Therefore, the potential energy increases at the expense of the internal energy of the fluid. Since APE remains close to zero, TPE $\approx$ BPE, and $\phi_d (0.001) \approx \phi_i (0.0008)$ from equation \ref{dAPEdt} till $\omega t \simeq 280$. This is demonstrated in figure \ref{subfig:F1fw59}. Notice that higher rotation rate suppresses turbulence (see Movie $3$), resulting in smaller $t.k.e$ and $B_V$ (figure \ref{subfig:F1fw59}) compared to the previous cases. $B_V$ increases after $\omega t \simeq 280$, increasing TPE. 
However, smaller $B_V$ signifies that only a minor portion of the TPE is available for conversion to $t.k.e.$ as shown by small APE in the inset of figure \ref{subfig:F1fw59}. The bulk of TPE goes in BPE, and the evolution of $\phi_d$ shows its rate of increase. The instability never saturates for F1f/$\omega$59 and triggers continuously, resulting in the sustenance of $t.k.e.$ and $B_V$. The TPE also continues to increase due to $B_V$, and APE continues to remain small. The $\phi_d$ oscillates and remains greater than zero indicating a continuous increase in BPE with time and signifying the sustenance of irreversible mixing for higher rotation rates. 
For $F = 2, 3$ and $f/\omega = 0, 0.48$ the $t.k.e.$, $B_V$, TPE, and APE increase from the beginning of the periodic forcing (figure not shown). Subsequently, BPE and $\phi_d$ also increase. Since the instabilities saturate quickly for these cases, $B_V$ and APE become $\sim$ zero. The BPE saturates, and the $\phi_d$ also tends to zero, signifying the ceasing of irreversible mixing. The initial evolution of energetics for $F = 2,3$ for $f/\omega = 0.59$ (figure \ref{subfig:F3fw59} for F3f/$\omega$59) is similar to the cases discussed above. However, at a later stage, the evolution resembles the case F1f/$\omega$59 and demonstrates non-zero $B_V$ and $\phi_d$, signifying a continuous increase in BPE and sustenance of irreversible mixing. We summarize the energy pathways in figure \ref{subfig:BPE}.\\
  \captionsetup[subfigure]{textfont=normalfont,singlelinecheck=off,justification=raggedright}
The time evolution of  $\eta_{cu}$ is plotted in the figure \ref{subfig:eff}. For F075f/$\omega$0 case, $\eta_{cu}$ decreases rapidly when the diapycnal flux $\phi_d$ and viscous dissipation $\epsilon_V$ increases. It attains a minimum value of $\sim 0.52$ after instability saturation. We find that in the presence of rotation for F075f/$\omega$48, the $\eta_{cu}$ reduces to $\sim 0.42$, which implies that the mixing is not as efficient as for F075f/$\omega$0. For F075f/$\omega$59, the $\eta_{cu}$ decreases continuously owing to the continuous irreversible mixing. However, the mixing is still more efficient ($\eta_{cu}\approx 0.57$ at $\omega t\simeq950$) than F075f/$\omega$0 and F075f/$\omega$48. For F1f/$\omega$0, $\eta_{cu}$ is $\sim 0.44$ similar to that obtained by \citet{briard2019harmonic}. At $F=1$ and $f/\omega = 0.48$ $\eta_{Cu} \sim 0.4$ whereas for $f/\omega = 0.59$ the cumulative mixing efficiency is 0.433 at $\omega t = 600$, but keeps decreasing with time. We find that the mixing becomes less efficient with an increase in $F$ for all $f/\omega$ values, and the mixing efficiency continues to decrease for the $f/\omega=0.59$ cases.
\section{Conclusions}\label{sec:conclusions}
We perform DNS to investigate the effect of rotation on the turbulent mixing driven by Faraday instability in two miscible fluids subjected to vertical oscillations. At lower forcing amplitudes $F=0.75$ and $1$, the $t.k.e.$ increases with an increase in the Coriolis frequency from $f/\omega=0$ to $0.48$. This enhancement is due to the excitement of more unstable $\theta$-modes in the sub-harmonic region resulting in the continuous development and breaking of KH billows at the interface layer during the sub-harmonic instability phase. The mechanism of turbulence generation for the cases without rotation is different. A well-defined mushroom-shaped wave evolves for $f/\omega=0$  cases that disintegrate at the nodes. Subsequently, the turbulence spreads to the entire mixing layer in successive oscillations. In contrast to $f/\omega=0, 0.48$, $f/\omega=0.59$ significantly suppresses turbulence owing to the development and gradual breaking of finger-shaped structures. At higher forcing amplitudes $F=2$ and $3$ for $f/\omega=0$, $0.48$, and $0.59$, we observed the breakdown of finger-shaped structures in the first two oscillations of periodic forcing resulting in the early onset of turbulence. However, the turbulence is less intense and short-lived than at lower forcing amplitudes owing to the shorter sub-harmonic instability phase. An interesting finding is an increase in $t.k.e.$ after an initial decrease for $f/\omega=0.59$ due to the continuous triggering of the sub-harmonic instabilities.\\ 
We also analyze the energetics associated with the flow. An increase in $t.k.e.$ for the cases with $f/\omega=0.48$ at $F=0.75,1$ implies an increase in the buoyancy flux and therefore, TPE also increases. A portion of this TPE is APE. Some part of this APE converts back to $t.k.e.$ via the buoyancy flux, whereas the rest is converted to internal energy, increasing the BPE through $\phi_i$. The rest of the TPE goes to BPE through $\phi_d$ resulting in irreversible mixing. When the $t.k.e.$ starts decreasing due to instability saturation, $B_V$ also decreases, signifying a decrease in APE. Therefore, BPE saturates, and $\phi_d$ approaches zero. Interesting, for $f/\omega=0.48$ the irreversible mixing sustains for a longer period than $f/\omega=0$ due to a longer sub-harmonic instability phase. The cases with a higher rotation rate of $f/\omega=0.59$ demonstrate a significant delay in the onset of sub-harmonic instability, owing to which $B_V$ and APE remain zero. The initial concentration profile diffuses during this period to increase TPE at the rate $\phi_i$. This TPE goes completely to BPE at the same rate since $\phi_d$ becomes equal to $\phi_i$ according to equation \ref{dAPEdt}. After the onset of the sub-harmonic instability, the $t.k.e$, $B_V$, and  APE increase. However, these values remain significantly smaller than  $f/\omega=0, 0.48$ cases. The TPE also increases due to $B_V$. Since APE is small, the bulk of the portion of TPE expends to BPE via $\phi_d$. Since the instabilities never saturate for $f/\omega=0.59$, the $t.k.e.$, $B_V$ and APE although small remains non-zero. Therefore, TPE increases continuously and expends to BPE via $\phi_d$. The $\phi_d$ remains greater than zero, signifying a continuous increase in BPE and irreversible mixing. At $F=2,3$, $\phi_d$ increases from the beginning of the vertical forcing for all $f/\omega$ cases. However, the irreversible mixing for $f/\omega = 0, 0.48$ does not sustain for long owing to the quick saturation of the sub-harmonic instability. Since the instability never saturates for $f/\omega = 0.59$, $\phi_d$ demonstrates continues oscillations and therefore, continuous mixing. Finally, the mixing phenomenon is quantified using cumulative mixing efficiency. We conclude that for rotation rates $f^2/\omega^2 > 0.25$ and at lower forcing amplitudes $F = 0.75,1$ the mixing process is more efficient. \\   
\backsection[Supplementary data]{\label{SupMat} Supplementary movies are available at https://doi.org/**.****/jfm.***...}
\backsection[Acknowledgements]{We gratefully acknowledge Dr. Vamsi K. Challamalla for providing us the computer program for the calculation of TPE, $x_3^*$, BPE and $\phi_d$. 
}





\bibliographystyle{jfm}
\bibliography{jfm}






\end{document}